\documentclass[%
 aip,
rsi,%
 amsmath,amssymb,
 reprint,%
]{revtex4-1}
\usepackage{ulem}
\usepackage{graphicx}
\usepackage{dcolumn}
\usepackage{bm}

\usepackage[usenames]{color}
\usepackage{colortbl}
 
\begin{document}

\preprint{AIP/123-QED}

\title{Measurement of local optomechanical properties of a direct bandgap 2D semiconductor}


\author{F.~Benimetskiy}
\email{fedor.benimetskiy@metalab.ifmo.ru}
\affiliation{ Department of Physics and Engineering, ITMO University, St. Petersburg 197101, Russia}

\author{V.~Sharov}
\affiliation{Ioffe Physical-Technical Institute RAS, St. Petersburg 194021, Russia}%
\affiliation{Academic University, St. Petersburg 194021, Russia}%
\author{P.A.~Alekseev}
\affiliation{Ioffe Physical-Technical Institute RAS, St. Petersburg 194021, Russia}%

\author{V.~Kravtsov}%
\affiliation{ Department of Physics and Engineering, ITMO University, St. Petersburg 197101, Russia}%

\author{K.~Agapev}%
\affiliation{ Department of Physics and Engineering, ITMO University, St. Petersburg 197101, Russia}%

\author{I.~Sinev}%
\affiliation{ Department of Physics and Engineering, ITMO University, St. Petersburg 197101, Russia}%

\author{I.~Mukhin}%
\affiliation{Academic University, St. Petersburg 194021, Russia}%

\author{A.~Catanzaro}%
\affiliation{ Department of Physics and Astronomy, University of Sheffield, Sheffield S3 7RH, United Kingdom}%

\author{R.~Polozkov}%
\affiliation{ Department of Physics and Engineering, ITMO University, St. Petersburg 197101, Russia}%

\author{A.~Tartakovskii}%
\affiliation{ Department of Physics and Astronomy, University of Sheffield, Sheffield S3 7RH, United Kingdom}%

\author{A.~Samusev}%
\affiliation{ Department of Physics and Engineering, ITMO University, St. Petersburg 197101, Russia}%

\author{M.~S. Skolnick}
\affiliation{ Department of Physics and Astronomy, University of Sheffield, Sheffield S3 7RH, United Kingdom}
\affiliation{ Department of Physics and Engineering, ITMO University, St. Petersburg 197101, Russia}%

\author{D.~N. Krizhanovskii}%
\affiliation{ Department of Physics and Astronomy, University of Sheffield, Sheffield S3 7RH, United Kingdom}
\affiliation{ Department of Physics and Engineering, ITMO University, St. Petersburg 197101, Russia}%

\author{I.~A. Shelykh}%
\affiliation{ Department of Physics and Engineering, ITMO University, St. Petersburg 197101, Russia}
\affiliation{Science Institute, University of Iceland, Reykjavik  IS-107, Iceland}

\author{I.~Iorsh}%
\affiliation{ Department of Physics and Engineering, ITMO University, St. Petersburg 197101, Russia}%

\date{\today}

\begin{abstract}
Strain engineering is a powerful tool for tuning physical properties of 2D materials, including  monolayer transition metal dichalcogenides (TMD) -- direct bandgap semiconductors with strong excitonic response. 
Here, we demonstrate an approach for local characterization of strain-induced modification of excitonic photoluminescence in TMD-based materials. We reversibly stress a monolayer of MoSe$_2$ with an AFM tip and perform spatio-spectral mapping of the excitonic photoluminescence in the vicinity of the indentation point. To fully reproduce the experimental data, we introduce the linear dependence of the exciton energy and corresponding photoluminescence intensity on the induced strain. Careful account for the optical resolution allows extracting these quantities with good agreement with the previous measurements, which involved macroscopic sample deformation. Our approach is a powerful tool for the study of local optomechanical properties of 2D direct bandgap semiconductors with strong excitonic response. 

\end{abstract}

\keywords{Transition metals dichalcogenides, 2D materials, strain engineering, excitons}
\maketitle


In recent years, single-layer transition metal dichalcogenides, 2D direct bandgap semiconductors, have attracted focused attention due to their unique electronic and optical properties \cite{mak2016photonics}. Mechanical strain is an important degree of freedom for wide-range tuning of carrier mobility\cite{hosseini2015strain}, bandgap\cite {johari2012tuning}, exciton energy 
\cite {niehues2018strain} and other properties in TMDs \cite{deng2018strain}. This becomes possible since TMDs sustain mechanical strain as large as 10$\%$ without rupturing \cite{bertolazzi2011stretching}. Strain-induced effects in TMD materials have been comprehensively studied via  macroscopic bending, stretching, or compressing the hosting substrate\cite{deng2018strain}. 



Furthermore, the planar geometry of TMDs provides a unique opportunity to use local strain for creating single-photon emitters through 3D quantum confinement of carriers \cite{feng2012strain, kumar2015strain, shepard2017nanobubble}. Such artificial atoms can be precisely positioned and arranged in lattices by, e.g., transferring TMD on nano-patterned substrates\cite {li2015optoelectronic, branny2017deterministic}. 
Another way to realize single-photon emitters with highly reproducible properties is nanoindentation of a TMD monolayer deposited on a deformable polymer substrate with atomic force microscope (AFM) \cite{rosenberger2019quantum}. 

One of the most straightforward and commonly used ways to study and visualize strain-induced spatial modulation of optical properties in a TMD monolayer is via confocal spectral mapping of the excitonic photoluminescence (PL) signal \cite{liu2014strain, kumar2015strain, li2015optoelectronic, branny2017deterministic, kumar2018nanosculpting}. It should be noted, however, that the scale of the deformation profile induced by substrate pattern or AFM nanoindentaion is often deeply subwavelength\cite{park2016hybrid}, which severely limits the precision of any direct estimations of the optomechanical properties of TMDs from far-field optical measurements\cite{kumar2015strain}. 

In this work, we use reversible AFM nanoindentation combined with PL mapping to study the strain-induced modification of the local exciton energy profile in a MoSe$_2$ monolayer. 
We show that careful account for the spatial resolution of the optical setup allows to extract with high precision the value of deformation potential, that may otherwise be strongly underestimated. Our approach and reconstruction routine is useful for non-destructive in-situ characterization\cite{li2018mechanical} of local optomechanical properties of TMD-based structures.

\begin{figure}[!ht]
\begin{minipage}[h]{1\linewidth}
\center{\includegraphics[width=1\linewidth]{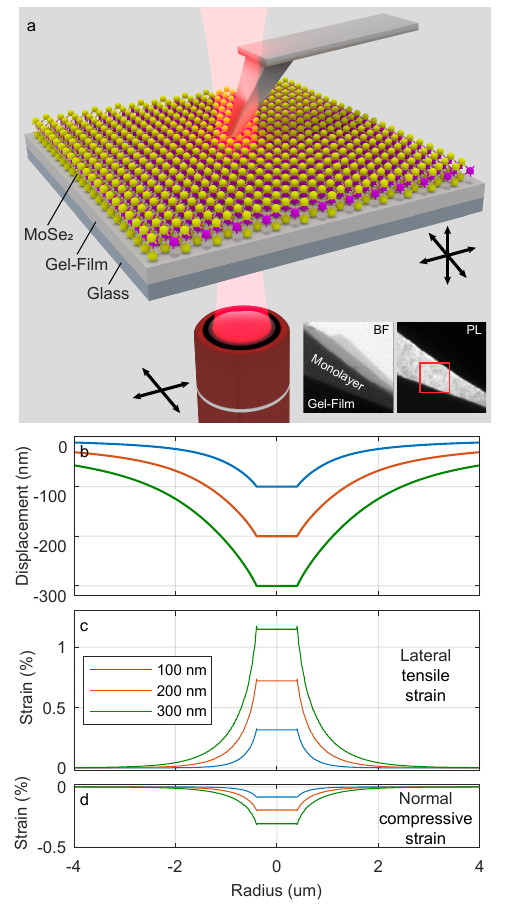}}
\end{minipage}

\caption{\label{fig1}  (a) Schematic of the experimental setup for local strain engineering in TMDs with atomic force microscopy tip. The inset shows bright-field (BF) and photoluminescence (PL) images of a part of the sample. The red square marks the region that was mapped in further experiments. [(b)-(d)] Calculated distribution of (b) MoSe$_2$ displacement, (c) lateral  and (d) normal components of strain for different (AFM tip) indentation depths.}
\end{figure}



MoSe$_2$ monolayer samples were fabricated by mechanical exfoliation with adhesive tape from a bulk crystal onto a polymer film (Gel-Film\textsuperscript{\textregistered} WF x4 6.0~mil) on SiO$_2$ substrates (Fig.~\ref{fig1}(a)). Monolayers were identified with fluorescence microscopy under 405~nm laser diode illumination through characteristic PL signal increase\cite{Tonndorf:13}. Insets in Fig.~\ref{fig1}(a) show bright field (left) and PL (right) images of a selected monolayer sample used in our experiment. 

\begin{figure*}[ht]
\begin{minipage}[h]{1\linewidth}
\center{\includegraphics[width=1\linewidth]{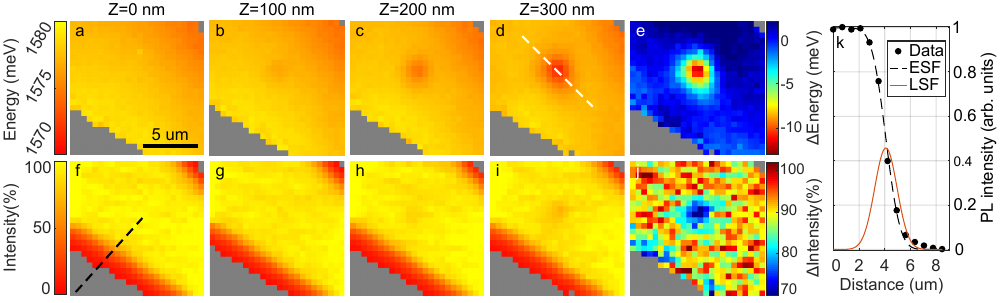} }
\end{minipage}
\caption{\label{fig2} Maps of the MoSe$_2$ monolayer PL peak intensities (a-d) and peak energies [(f)-(i)] for different indentation depths (0, 100, 200 and 300 nm). [(e) and (j)] represent the respective differential quantities for 300 nm indentation. (k) PL intensity profile across the MoSe$_{2}$ edge as illustrated by dashed line in panel (f). The experimental data are shown with dots. Dashed and solid line represent the edge spread function (ESF) and line spread function (LSF) fits, respectively. The waist of the error function is equal to 1.69  $\mu$m as determined by the ESF fit.}
\end{figure*}

The experimental setup is sketched 
in Fig.~\ref{fig1}. We employ 
indentation by an AFM tip to investigate the fundamental optomechanical properties of locally strained TMD flakes. The curvature radius of the Si cantilever tip that we used for indentation was approximately 400 nm. The tip was intentionally blunted by focused ion beam (FIB) milling to avoid rupturing of the flake and increase the achievable strain threshold. After the milling procedure the tip had a plane facet orthogonal to the indentation direction (\textit{i.e.} parallel to MoSe$_2$ surface). The indentation was performed using AIST SmartSPM\textsuperscript{TM} module in contact mode with a stiff probe (NT-MDT VIT$\_$P, 50~N/m) to ensure that the cantilever bending is negligible in comparison to the sample deformation.
Taking into account the very high elasticity of Gel Film (the film is a polysiloxane-based polymer similar to poly-dimethyl siloxane\cite{Castellanos-Gomez14}), the vertical displacement of the sample relative to the stationary cantilever from the tip-sample contact position was directly interpreted as the indentation depth. We further confirmed the absence of cantilever bending by measuring force--distance curves.

To determine the strain of the MoS$_2$ monolayer induced by the local deformation with AFM tip we used COMSOL Multiphysics.
For simplicity, cylindrical symmetry of our model was assumed. In particular, the MoSe$_2$ flake was represented by a membrane with a radius of 75~$\mu m$
and thickness of 0.7 nm perfectly bonded to GelFilm substrate with a radius of 100 um. Young's moduli for MoSe$_2$ and GelFilm were chosen as 178 GPa and 0.5 MPa, respectively~\cite{yang2017brittle,iguiniz2019revisiting}.
The displacement and strain distribution profiles calculated for different AFM tip indentation depths are presented in [Fig.~\ref{fig1}(b)-\ref{fig1}(d)].

As seen in Fig.~\ref{fig1}(b), the deformed region of the MoSe$_2$ flake is significantly larger than the tip--flake contact region. This is due to very large (10$^5$) difference in Young's moduli of MoSe$_2$ and GelFilm and agrees well with the recent results by Niu \textit{et al.} \cite{NIU20181} showing that indentation of a stiff 2D membrane on the compliant substrate leads to an increase of the deformed area with respect to the AFM tip diameter. 
Furthermore, the modelling of the deformation process in COMSOL shows that the lateral tensile strain of the membrane (1.2\% beneath the tip  for indentation depth of 300 nm, Fig.~\ref{fig1}(c)) is approximately four times larger than the normal compressive strain (0.35\%, Fig.~\ref{fig1}(d)).

In order to perform in-situ optical characterization of the induced strain profile, we have carried out spectral mapping of the PL signal for different indentation depths. In the experiment, the sample was excited from the substrate side with 632.8 nm HeNe CW laser focused with a 100х/0.7 objective lens (Mitutoyo M Plan Apo). PL signal was collected with the same objective and analyzed by Horiba LabRAM spectrometer in the confocal arrangement.  
To map the spatial distribution of deformation-driven PL spectral shift, we used a piezo stage mounted on the objective holder, so that during the scan the probe and the sample remained stationary.

\begin{figure}[ht]
\begin{minipage}[h]{1\linewidth}
\center{\includegraphics[width=1\linewidth]{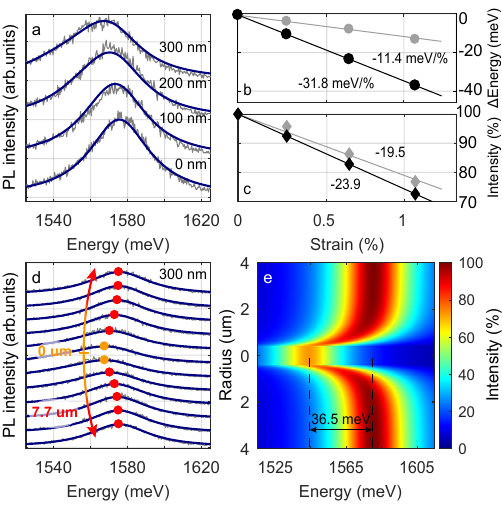} }
\end{minipage}
\caption{\label{fig3} (a) PL spectra at the center of deformation for different indentation depths. Peaks are fitted with the proposed model. The strain dependence of (b)  the exciton peak and (c) PL intensity. Black solid lines and markers in [(b) and (c)] correspond to the extracted parameters. Grey markers and lines correspond to underestimated values obtained directly from the PL spectra measured at the indentation point.  (d) PL spectra measured across the deformation area as illustrated by the dashed line in Fig.~\ref{fig2}(d). Spectra at the center of local deformation are labeled by orange symbols. (e) Strain profile dependence of the modelled spectra for the case of 300 nm indentation. }
\end{figure}

PL peak position and intensity maps obtained from a MoSe$_2$ flake for different indentation depths are shown in [Fig.~\ref{fig2}(a)-\ref{fig2}(d)] and [Fig.~\ref{fig2}(f)-\ref{fig2}(i)], respectively. The peak position maps reveal minor inherent heterogeneity, which can be attributed to the initial tension accumulated during flake transfer on GelFilm. This is confirmed by corresponding differential maps for PL peak position shift (Fig.~\ref{fig2}(e)) and intensity decrease (Fig.~\ref{fig2}(e)) induced by tip indentation.
Upon the increase of the indentation depth, the area of local deformation becomes apparent from the PL intensity maps (Fig.~\ref{fig2}(a)-\ref{fig2}(d)). At the point of tip impact, the observed PL peak weakens and exhibits a red-shift, as shown in Fig.~\ref{fig3}(b) and \ref{fig3}(c) (grey markers), which is consistent with previous works\cite{Island2016,liang2017monitoring}. In the experiment, the maximum measured displacement of the PL peak was $\sim 13$ meV and the PL intensity at the indentation point decreased by 22.3$\%$.  

To check how the measured results correspond to the local strain-induced modification of the optical properties of TMDs, we have carried out density functional theory (DFT) calculations. Theoretical investigation of electronic and band structures allows estimating strain-dependent shifts of the exciton spectral peak position. For these computations, we use generalized gradient approximation (GGA) \cite{Kurth99, Staroverov04} that is local but takes into account the gradient of electron density as a correction, with Perdew, Burke and Ernzerhof (PBE) functionals\cite{Perdew96, Burke98}. 

It is well known that the standard DFT approach underestimates the band gap of the material \cite{Perdew85}. One of the best ways to overcome this disadvantage is employing the $GW$ approach \cite{Aryasetiawan98, Hedin65, Held11} that allows us to expand the self-energy in terms of the single particle Green's function and screened Coulomb potential. From this point of view, it is convenient to use  $G_0W_0$ approximation to obtain realistic values of the band gap energy.

The exciton peak energy was obtained by solving Bethe-Salpeter equation (BSE) \cite{BSE, Benedict99}, with occupied valence and unoccupied conduction band states with the energies obtained from the DFT and DFT with electronic band structure that was corrected by $G_0W_0$ quasi-particle energies. The Bethe-Salpeter equation is a way to improve the system description, that describes the bound states of a two-body (particles) and interaction of such bound states with each other. It is very important for monolayer calculation to consider two quasi-particles due to the weak screening and, as a result, strong excitonic effect. Therefore, Lanczos-Haydock algorithm \cite {Haydock80} combined with Tamm-Dancoff approximation \cite {Tamm45, Dancoff50} was used to solve BSE.

Our first-principles DFT calculations were performed using Quantum Espresso package\cite{Giannozzi09, Giannozzi17}. $G_0W_0$ correction and the exciton effects were calculated via Yambo project package \cite{Marini09}.

As a result of these ab-initio calculations, we obtain the deformation potential for exciton energy of $-43.5$ meV/\% and 10 meV/\% for lateral stretching and normal compression, respectively. The difference between DFT+BSE and GW+BSE results is negligible. 
Note that according to the simulation results, the deformation potential attributed to normal strain is approximately four times smaller than that corresponding to lateral strain. Taking into account the fact that in our experimental geometry the normal strain is four times weaker than the lateral one, its contribution will be neglected in further considerations. 



The exciton peak shift obtained directly from PL mapping at 300~nm indentation (Fig.~\ref{fig2}(e)) is about four times smaller than that predicted theoretically by ab-initio calculations combined with simulated strain distribution (see Fig~\ref{fig1}(c)). This significant discrepancy is due to the limited spatial resolution of the optical setup. In the experiment, the diameter of the effective collection area (waist $w$) can be estimated from the PL intensity profile along the direction perpendicular to the flake edge (see Fig.~\ref{fig2}(k)). The measured profile is best fitted by a function $1 - erf(2 - \sqrt(x)/w)$, with $w = 1.69~\mu$m. Therefore, the resolution is of the order of the spatial scale of the deformation, which can result in discrepancies with the theoretical predictions. 

In order to account for the optical resolution, fully reproduce the experimental data and extract the optomechanical properties of a TMD monolayer, we first introduce a simple model, which relates the local Lorentzian-shaped PL spectrum with strain in each spatial location. In the linear approximation, we assume that the induced strain results in 
linear exciton energy shift and linear PL intensity change 
with respective coefficients $\xi$ and $\eta$.
We also assume that the spectral width of the exciton PL remains unchanged, which is adequate in the first-order approximation\cite{niehues2018strain}.
By performing the spatial convolution of the local spectra calculated for given $\xi$ and $\eta$ with the Gaussian point spread function (PSF, see Fig.~\ref{fig2}(k)), one can simulate the spectral maps expected in the experiment. 

Fig.~\ref{fig3}(a) and \ref{fig3}(d) shows measured (grey lines) and simulated (blue lines) spectra obtained with the optimized parameters $\xi = -31.8$ meV/$\%$ and $\eta = 23.9$. Their direct comparison confirms very good agreement between theory and experiment. Moreover, the obtained optomechanical parameters are well-correlated with our ab-initio simulations and previously reported results for MoSe$_2$\cite{niehues2018strain, Island2016}. 
This readily supports the feasibility of the introduced model for the relation between strain and excitonic PL spectrum. Fig.~\ref{fig3}(b) clearly shows that in our experimental geometry deformation potential $\xi$ extracted with account for the optical resolution is three times larger than that obtained directly from the measured PL spectrum at the indentation point. Meanwhile, the factor $\eta$ responsible for the strain-induced change of PL intensity appears less sensitive to the resolution (see Fig.~\ref{fig3}(c)). Finally, we plot radial dependence of local exciton PL spectra corresponding to maximum indentation of $Z = 300$~nm (Fig.~\ref{fig3}(e)). This map demonstrates that the achieved 
exciton energy shift of 36.5~meV is of the order of PL half-width of 38 meV
and less than thermal energy of 25.7~meV under ambient conditions. This result paves the way towards development of single-photon sources with controllable properties by means of local elastic and reversible indentation at elevated temperatures.

To conclude, we propose an approach allowing to perform in-situ characterization of local optomecanical properties of a TMD monolayer. In the experiment, we expose the MoSe$_2$ flake to a local indentation by an AFM tip. For each indentation depth, we carry out spectral mapping of the PL signal. In order to reproduce the experimental data, we introduce linear relation of TMD strain with the parameters of the Lorentzian-shaped exciton PL spectrum: the shift of exciton energy and the change in the PL intensity. Using the simulated spatial distribution of strain induced by an AFM tip with the careful account for the optical resolution, we fully reproduce the whole set of the experimental data and extract the optomechanical properties of the TMD monolayer. The obtained values are in good agreement with the previous measurements based on macroscopic strain and the results of our ab-initio simulations. This opens new ways for the studies of local optomechanical properties of 2D direct-bandgap semiconductors. Furthermore, the exciton energy profile obtained under maximum elastic deformation supports the fact that reversible local indentation may be used for controllable engineering of single-photon sources at elevated temperatures.

The authors acknowledge funding from the Ministry of Education and Science of the Russian Federation (Megagrant No. 14.Y26.31.0015, Zadanie No.3.8891.2017/8.9, Zadanie No. 3.1365.2017/4.6).  V.K. acknowledges financial support from ITMO Fellowship.

\bibliography{aipsamp}
\end{document}